\documentclass[epj]{webofc}
\usepackage[varg]{txfonts}   
 \newcommand{\nn}{\nonumber}
 \newcommand{\sura}{\hspace{-1.5mm}\!/}

\woctitle{QCD@Work 2014}
\begin{document}
\title{Energy profile of b-jet for boosted top quarks}
%
%


\author{Yoshio Kitadono\inst{1}\fnsep\thanks{\email{kitadono@phys.sinica.edu.tw}} 
}

\institute{
Institute of Physics, Academia Sinica, Taipei 11529, Taiwan, Republic of China
          }

\abstract{%
  We analyse the semileptonic decay of a polarised top-quark with a large velocity based on the perturbative QCD factorisation framework. Thanks to the factorisation and the spin decomposition, the production part and the decay part can be factorised and the spin dependence is introduced in the decay part. The decay part is converted to the top-jet function which describes the distribution of jet observables and the spin is translated to the helicity of the boosted top. Using this top-jet function, the energy profile of b-jet is investigated and it is turned out that the sub-jet energy for the helicity-minus top is accumulated faster than that for the helicity-plus top. This behaviour for the boosted top can be understood with the negative spin-analysing-power of b-quark in the polarised-top decay.
}
\maketitle
\section{Introduction}\label{intro}
Top-quark polarisation is important to reveal a hint of beyond the standard model, especially chiral structures of the top sector. The information of the polarisation for a highly boosted top-quark is translated into that of helicities, since the chirality coincides with the helicity at high energies. 
Once the boosted top-quark is produced, the top quark decays through the weak interaction and the decay particles of the highly boosted top will be collimated along the direction of the top momentum. Such highly collimated configurations will spoil ordinary methods to distinguish these decay particles. Jet substructures are useful to overcome this difficulty in jets from boosted particles. Various jet observables are proposed, for example, "girth" \cite{girth} and "angularity" \cite{angularity} are useful to discuss sub structures of jets.  For the readers who are interested in the jet substructure, see the recent review (lecture note) \cite{review.jetsubstructure} for the jet substructure. 

As the top-polarisation effects can be found in the sub-jet-energy distribution \cite{boost.top.pol}, the research in a correlation between top polarisation and other jet-substructures will be important to study not only for a deeper understanding the standard model, but also to discuss an extended chiral structure in the top sector.
In our previous work \cite{YKLi}, the helicity dependence in the top-jet substructure is discussed, especially the energy profile (alternatively, jet shape) which expresses a distribution of the sub-jet energy in the top-jet cone for a highly boosted polarised-top is considered. The energy profile is defined in the following:
\begin{eqnarray}
\Psi(r) &=&
\frac{1}{N_{J_t}}\sum_{J_t}\frac{\sum_{r_i<r, i\in J_t}P_{T_i}}
{\sum_{r_i<R_t, i\in J_t}P_{T_i}},\label{profile}
\end{eqnarray}
where $r \le R_t$ is a test radius in the top jet $J_t$, 
$N_{J_t}$ is a number of top jets with the top-jet radius $R_t$, $P_{T_i}$ is the transverse momentum of a particle $i$ in the top-jet. It is worthy to note that the lepton energy in the semileptonic top-decay is not included in the definition. 

In our framework, this energy profile is expressed as a convolution of a hard kernel with the energy function for the light-quark-jet evaluated by pQCD calculation. By using this energy profile for the top-jet with a particular helicity, we can theoretically study the distribution of the sub-jet energy in the top decay, especially, we consider the semileptonic decay for simplicity of the analysis and we count the energy of b-jet in this decay. We will discuss the top-jet-energy dependence and the helicity dependence in the energy profile. The helicity-dependence is converted into a "helicity minus-plus (chirality left-right) difference" and this difference will be useful to distinguish the helicities of top quark. It is turned out that the energy profile is sensitive to the helicities of the top quark and the top quark with helicity-minus can accumulate the sub-jet energy faster than one with helicity-plus. This feature is understood within the standard model, namely within the $V-A$ structure of the weak interaction. Theoretical formalism and the results of the energy profile are presented in the section \ref{formalism}, the reason of the difference in the energy profile between different helicities is discussed in the section \ref{discussion}, and the section \ref{conclusion} is devoted to the conclusion.



\section{Factorisation and Energy profile}\label{formalism}

\subsection{formalism}
We consider the process $q\bar{q} \to t\bar{t}$ as the subprocess of the top-pair production to construct the top-jet function $J_t$. In principle, we can include the subprocess $gg \to t\bar{t}$, but the factorisation procedure for $gg$ process is common as well as $q\bar{q}$. Therefore we only consider $q\bar{q} \to t\bar{t}$ process as the production process for simplicity.
The factorisation at the leading order (LO) is simple. We can change the fermion flow thanks to the Fierz identity and the production part and the decay part are factorised in the squared matrix-element as
\begin{eqnarray}
  \Big| \overline{\mathcal{M}} \Big|^2
&=&
 \Big| \overline{\mathcal{M}}_{pro} \Big|^2
 \Big| \overline{\mathcal{M}}_{decay} \Big|^2
\left[1 + O\left(\frac{m^2_t}{s}\right)\right],
\end{eqnarray} 
where $\overline{\mathcal{M}}$ is the total probability-amplitude of the process $q\bar{q} \to \bar{t} b\ell \nu$, $\overline{\mathcal{M}}_{pro}$ is the production part related to the process $q\bar{q} \to t\bar{t}$, $\overline{\mathcal{M}}_{decay}$ is the decay part related to the process $t\to b\ell \nu$, $\sqrt{s}$ is the centre of mass energy for $q\bar{q}$ pair. We can neglect the term $\mathcal{O}(m^2_t/s)$ for a highly boosted top-quark. The production part is canceled out in the final result of the energy profile, hence we don't explicitly write the full expression. The decay part $\Big| \overline{\mathcal{M}}_{decay} \Big|^2$ is given by the product of leptonic trace and the trace for the decay part. Factorising the b-quark trace from the decay trace by Fierz identity and combining the phase space, the decay part is converted into a part of the top-jet function \cite{YKLi}.
The LO top-jet function $J^{(0),s_t}_{t}$ specified by the top-spin vector $s_t$ is expressed as the convolution with the hard kernels $F_a, F_b$ and the LO b-jet function $J^{(0)}_{b}$ in the following form:
\begin{eqnarray}
 J^{(0),s_t}_{t}(m^2_{J_t},\bar E_{J_t},\bar R_t)
 &=& f_{t}(z_{J_t}) \int dz_{J_b}d\bar x_{J_b}d\cos\bar\theta_{J_b}\nn\\
 &{}& \times
\left[  F_{a}(z_{J_t}, \bar x_{J_b}, z_{J_b}) + |\vec{s}_t|
F_{b}(z_{J_t}, \bar x_{J_b}, z_{J_b}) \cos\bar\theta_{J_b}
\right] J^{(0)}_{b}(m^2_{J_b}, \bar E_{J_b}, \bar R_t),\label{eq.Jtst.LO}
\end{eqnarray}
where $\bar E_{J_t}=m_{J_t}$, $\bar E_{J_b}$ is the b-jet energy in the
rest frame of the top quark, the dimensionless parameters $z_{J_t}$, $\bar{x}_{J_b}$, $z_{J_b}$ are defined as
\begin{eqnarray}
z_{J_t} = \frac{m^2_{J_t}}{m^2_t}, \hspace{1.0cm}
\bar x_{J_b} = \frac{2\bar{E}_{J_b}}{m_{J_t}}, \hspace{1.0cm}
z_{J_b} = \frac{m^2_{J_b}}{m^2_{J_t}}, 
\end{eqnarray}
and the polar angle $\bar\theta_{J_b}$ is measured as the relative angle between the top-spin $\vec{s}_t$ and the b-jet momentum, $\bar R_t$ is the top-jet radius supposed to be the upper bound of $\bar\theta_{J_b}$ in the rest frame of the top quark. The hard kernels $F_a$ and $F_b$ are given by
\begin{eqnarray}
 F_a(z_{J_t}, \bar{x}_{J_b}, z_{J_b}) &=&
 \sqrt{z_{J_t}} \sqrt{\bar{x}^2_{J_b} - 4z_{J_b}} f_{W}(z_{J_t},\bar{x}_{J_b}, z_{J_b})
       \left[   - \frac{1}{3}\bar{x}^2_{J_b} 
                + \frac{1 + z_{J_b}}{2}\bar{x}_{J_b}
                - \frac{2}{3} z_{J_b}
       \right],\nn\\
 F_b(z_{J_t}, \bar{x}_{J_b}, z_{J_b}) &=&
 f_{W}(z_{J_t}, \bar{x}_{J_b}, z_{J_b})
       \left[   - \frac{1}{3}\bar{x}^3_{J_b} 
                + \frac{1 + 3z_{J_b}}{6}\bar{x}^2_{J_b}
                + \frac{4}{3} z_{J_b} \bar{x}_{J_b}
                - \frac{2}{3} z_{J_b} (1 + 3z_{J_b})
       \right],
\end{eqnarray}
where $f_{W}(z_{J_t}, \bar{x}_{J_b}, z_{J_b})=1/[(1+z_{J_b}-\bar{x}_{J_b}-\xi)^2 + (\xi\eta)^2]$ is the dimensionless $W$-boson propagator with the mass ratios $\xi=m^2_W/m^2_{J_t}, \eta=\Gamma_W/m_W$. The overall factor $f_{t}(z_{J_t})$ is proportional to the dimensionless top propagator $1/[(1-z_{J_t})^2 + \eta^2_{t}]$ with the mass ratio $\eta_t=\Gamma_t/m_{J_t}$. The spin dependence in the top-jet function is introduced through the spin decomposition $(k_t\sura + m_t)=(k_t\sura+m_t)(1+\gamma^5 s_t\sura)/2+(k_t\sura+m_t)(1-\gamma^5s_t\sura)/2$.

Although the LO b-jet function $J^{(0)}(m^2_{J_b}, \bar{E_{J_b}}, R_{b})$ is proportional to the delta function $\delta(m^2_{J_b} - m^2_b)$, by taking into account the soft-gluon contribution to this process, we obtain the expression of the top-jet function $J^{s_t}_t$ including the QCD effects in the following form:
\begin{eqnarray}
J^{s_t}_{t}(m^2_{J_t},\bar E_{J_t},\bar R_t)
 &=& f_{t}(z_{J_t})\int dz_{J_b}d\bar x_{J_b}d\cos\bar\theta_{J_b}\nn\\
 &{}& \times
\left[  F_{a}(z_{J_t}, \bar x_{J_b}, z_{J_b}) + |\vec{s}_t|
F_{b}(z_{J_t}, \bar x_{J_b}, z_{J_b}) \cos\bar\theta_{J_b}
\right] J_{b}(m^2_{J_b}, \bar E_{J_b}, \bar R_t),\label{eq.Jtst.resum}
\end{eqnarray}
where the bottom-jet function $J_{b}(m^2_{J_b}, \bar E_{J_b}, \bar R_t)$ improved by the soft-gluon resummation is available. for instance, in Ref. \cite{energyprofile}.

In order to convert the rest frame of the top quark to its boost frame, we relate the top-jet energy $E_{J_t}$ and the decay angle $\theta_{J_b}$ of the b-jet  defined at the boost frame to those in the rest frame through the Lorentz transformation \cite{Shelton}
\begin{eqnarray}
 E_{J_t} = \gamma_t \bar{E_{J_t}}, \hspace{1.0cm}
 \cos\bar{\theta}_{J_b} = \frac{-v_t + \cos\theta_{J_b}}{1 - v_t \cos\theta_{J_b}} 	
\end{eqnarray}
where we neglect the b-jet mass, because it is smaller than the mass scale of the top-jet energy or top-jet mass. We can neglect $z_{J_b}$ dependent terms in the hard kernels $F_a, F_b$ due to the same reason. Here the Lorentz transformation is performed so that the momentum of the boosted top is along the spin direction of the top quark. Therefore we regard $J^{s_t}_{t}$ as the top-jet function $J^{R}_t$ with the helicity-plus (alternatively right-hand top)  by the above Lorentz boost. On the other hand, the top jet function $J^{L}_t$ with the helicity-minus (left-hand top) is expressed by replacing the sign of the cosine dependent term as $J^{L}_t = J^{R}_t \big|_{\cos \to -\cos}$.

The top-jet-energy function $J^{E,R(L)}_{t}$ for the right(left)-hand top is defined by the similar way as well as the top-jet function $J^{R(L)}_{t}$. Multiplying the transverse energy of the b-jet within a test cone $r<R_t$ to the integrand of the top-jet function and integrating out the top-jet mass, we can derive the top-jet-energy function $J^{E,R(L)}_{t}$ in the following form:
\begin{eqnarray}
J^{R(L)}_{t}(\bar E_{J_t},\bar R_t, r)
 &=& \int \frac{dz_{J_t}}{z_{J_t}}f_{t}(z_{J_t})
     \int dz_{J_b}d\bar x_{J_b}d\cos\bar\theta_{J_b}\nn\\
 &{}& \times
\left[  F_{a}(z_{J_t}, \bar x_{J_b}, z_{J_b}) \pm |\vec{s}_t|
F_{b}(\bar x_{J_b}, z_{J_b}) \cos\bar\theta_{J_b}
\right] J^{E}_{b}(\bar E_{J_b}, \bar R_t, r),\label{eq.JtEst.resum}
\end{eqnarray}
where the hard kernels $F_a, F_b$ are same functions appeared in $J^{R(L)}_t$, the energy function $J^{E}_{b}$ is calculated in Ref. \cite{energyprofile}.

\subsection{Results}
\begin{figure}
\centering
 \begin{tabular}{cc}
  \includegraphics[width=7.2cm,clip]{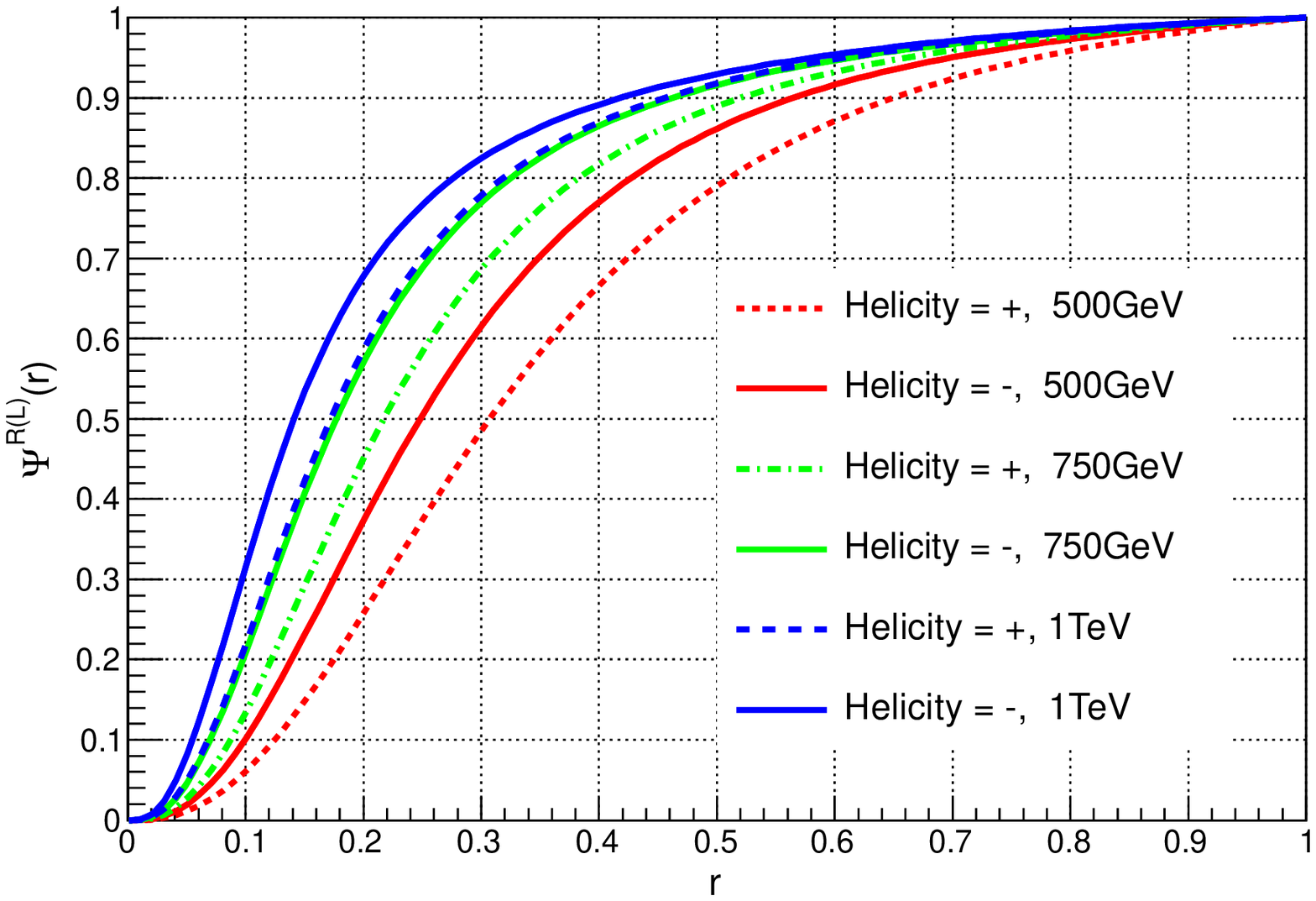} &
  \includegraphics[width=7.2cm,clip]{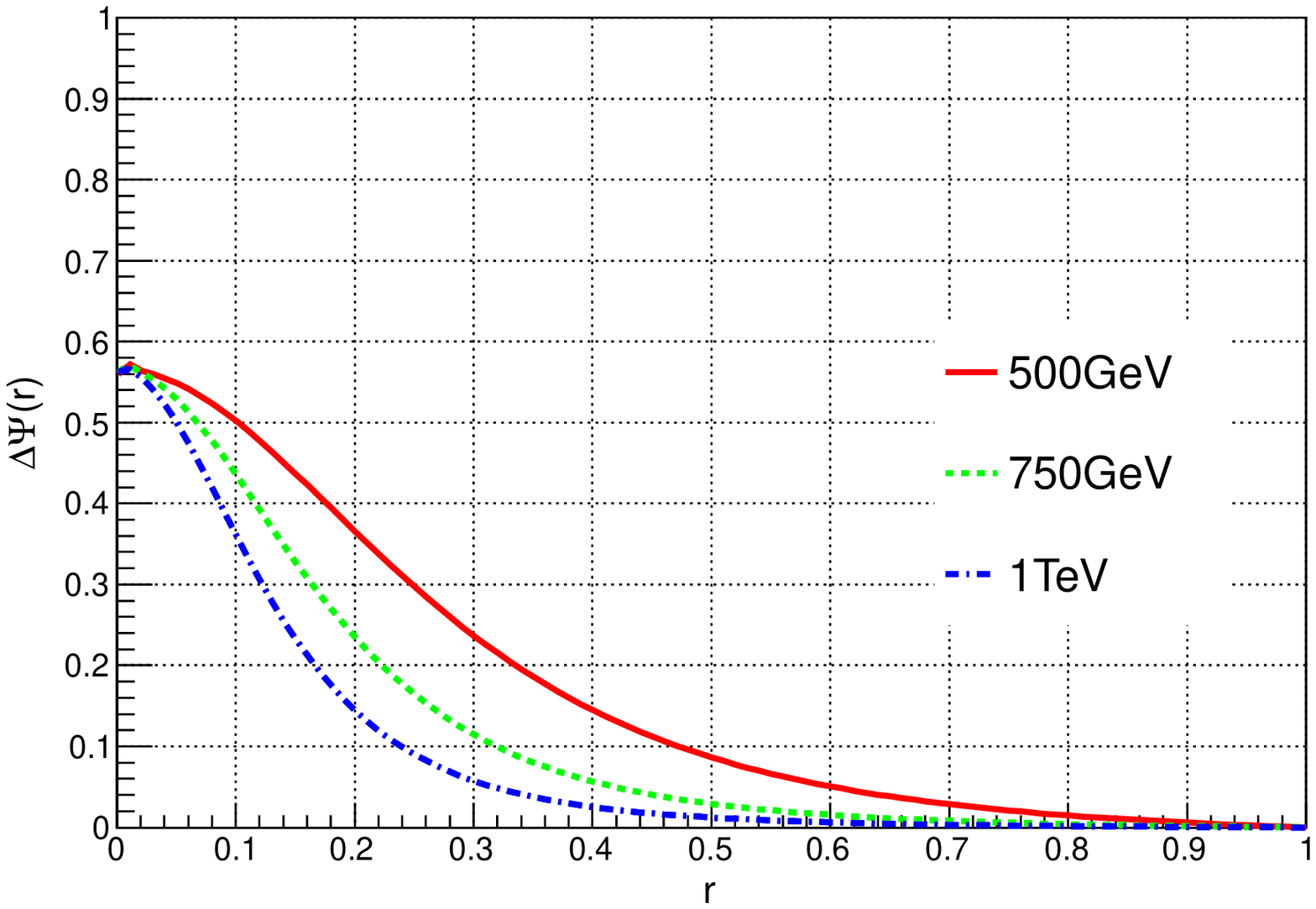} \\
   (a) &  (b)
 \end{tabular} 
\caption{Top-jet-energy $E_{J_t}$ dependence in the energy profile for the boosted top:(a), the helicity minus-plus (chirality left-right) difference $\Delta\Psi(r)$:(b).}
\label{fig-1}       
\end{figure}

The energy profile $\Psi^{R(L)}(r)$ at the parton level for the helicity-plus (minus) top is expressed in terms of the energy function $J^{E,R(L)}_t$ as the function of the test-cone radius $r$:
\begin{eqnarray}
 \Psi^{R(L)}(E_{J_t}, R_{t}, r) 
  = \frac{J^{E,R(L)}_{t}(E_{J_t},R_t, r)}
         {J^{E,R(L)}_{t}(E_{J_t},R_t, r=R_t)}.
\end{eqnarray}
The top-jet-energy $E_{J_t}~(\mbox{velocity}~\beta_t,~\mbox{gamma~factor}~\gamma_t)$ dependence in the energy profile is shown in (a) of Figure \ref{fig-1}.
We use the parameters $m_t=172.5~\mbox{GeV}, m_{W}=80.39~\mbox{GeV}$ for masses, $\Gamma_W=2.09~\mbox{GeV}, \Gamma_{t}=1.33~\mbox{GeV}$ for decay widths, and $\Lambda_{\mbox{\tiny QCD}}=0.1~\mbox{GeV}$ for the scale parameter of QCD with six flavours.

It is obvious that the energy profile of the helicity-minus (left-hand) top is larger than one of the helicity-plus (right-hand) top for $E_{J_t} = 500~\mbox{GeV}~(\beta_t=0.94, \gamma_t=2.9), 750 \mbox{GeV}~(\beta_t=0.97, \gamma_t=4.3)$ and $1~\mbox{TeV}~(\beta_t=0.99, \gamma_t=5.8)$ with a fixed top jet radius $R_t=1.0$. The difference of the energy profile between the helicity-plus top and the helicity-minus top can be evaluated with a difference between $\Psi^{L}(r)$ and $\Psi^{R}(r)$, for example with the value $\Delta \Psi(r)$ expressed by the following definition
\begin{eqnarray}
\Delta \Psi(r) = \frac{\Psi^{L}(r) - \Psi^{R}(r)}
                      {\frac{\Psi^{L}(r) + \Psi^{R}(r)}{2}},
\end{eqnarray}
where this value is the ratio of the difference between the helicity-minus and helicity-plus to its average. The ratio $\Delta\Psi(r)$ is shown in (b) of Figure \ref{fig-1}.
 Typical difference between the helicity-minus and helicity-plus top can be found at small $r$ region, for example, at $r=0.1$. The numerical values of $\Delta\Psi(r=0.1)$ are $50\verb|%|$, $43\verb|%|$, and $36\verb|%|$ for $E_{J_t}=500~\mbox{GeV}, 750~\mbox{GeV}, 1~\mbox{TeV}$ respectively. These differences decrease as the energy of the top increases as is expected by the kinematics of the special relativity. Although the differences are larger for lower energies of the top quark, it is important that the validity of our approximation used at the factorisation between the production part and the decay part is relatively lost at lower energies and the result for $E_{J_t}=1~\mbox{TeV}$ has the best validity among these three top-energies.

\section{Discussion}\label{discussion}
\begin{figure}

\centering
 \begin{tabular}{c}
  \includegraphics[width=12cm,clip]{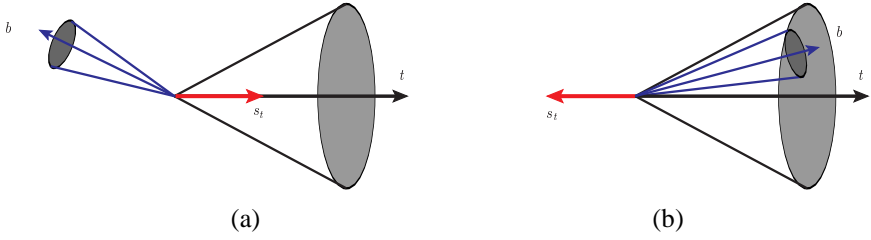}\\
  \hspace{0cm} (a) \hspace{5cm} (b)
 \end{tabular} 
\caption{Favoured decay direction of the b-jet for the top quark with the helicity-plus:(a) and helicity-minus:(b). }
\label{fig-2}       
\end{figure}

\noindent
The mechanism why the energy profile for the helicity-minus (left-hand) dominates than that of the helicity-plus (right-hand) is explained in Figure \ref{fig-2}. 
According to the standard $V-A$ weak interaction,  the angular distribution for decay particle of the polarised top-quark at the rest frame of the top is summarised in the following form \cite{Spin.analyz1,Spin.analyz2}:
\begin{eqnarray}
 \frac{1}{\Gamma}\frac{d\Gamma}{d\cos\theta_{i}}
  &=& \frac{1}{2}(1 + \kappa_i |\vec{\rho}| \cos\theta_{i}), \hspace{1cm}i=b,\ell,\nu,
\end{eqnarray}
where $\Gamma$ is the partial decay width of this decay process, $|\vec{\rho}|$ is the polarisation vector for the top, the decay angle $\theta_i$ is measured between the top spin and the momentum direction of the decay particle $i$, the numerical constant $\kappa_i$ is known as the spin-analysing-power which describes the sensitivity of the decay particle $i$ to the top spin. Numerically, the spin-analysing-power of b-quark is the negative value $\kappa_b \simeq -0.4$ \cite{Spin.analyz3} and it means that the favoured decay direction of the bottom quark is opposite to the top spin direction. 

This tendency will be kept as far as the boost is not too large, since the top quark is very heavy and the boost parameter is less than unity in actual experiments.
Hence there is a correlation between the momentum direction of the b-jet and the top helicity translated from the top spin. 
For example, according to the definition of the helicity, the b-jet tends to be emitted along the opposite direction to the top spin for the helicity-plus top  (right-hand top) and therefore the b-jet tends to go outside the top-jet cone ((a) of Figure \ref{fig-2}). On the other hand, the b-jet tends to be emitted along the same direction to the top spin for the helicity-minus top quark (left-hand top) and therefore the b-jet tends to go inside the top-jet cone ((b) of Figure \ref{fig-2}). 
Comparing the contribution to the jet energy profile both for the helicity-plus top and for helicity-minus top, the b-jet contribution to the energy profile for the helicity-minus top has a larger probability than one for helicity-plus top. This is the reason why the helicity-minus top can accumulate the b-jet energy faster than the helicity-plus top.

\section{Conclusion}\label{conclusion}
We have theoretically investigated the helicity dependence in the jet substructure within the standard model, especially, the energy profile of the top-jet for the semileptonic top-decay. The main result is expressed by the convolution with the b-jet energy function improved by pQCD resummation and the hard kernel calculated by the weak interaction. It indicates that the helicity-minus top can accumulate the energy of the b-jet in the semileptonic decay faster than the helicity-plus top.
This tendency can be understood with the standard $V-A$ weak interaction, i.e., the consequence of the negative spin-analysing-power for b-quark. These results imply that the energy profile, one of the simple jet-substructure, is useful for the discrimination of the helicities of the boosted top quark. 

Besides this discrimination will be helpful not only for the identification of the top helicities, but also for the study of the chiral structure of the top quark through jet observables. The straightforward application of this formalism to the hadronic top-decay is under investigation. We expect that a similar difference will appear in other observables of the jet substructure like girth \cite{girth} or angularity \cite{angularity} and other jet observables discussed in \cite{review.jetsubstructure}. 
Although the data of the energy profiles for light-jet and b-jet in top events at the LHC is reported in Ref. \cite{atlas.jetshape}, the comparison with our results and the experimental data is nontrivial, since we only focus on highly boosted top-quarks. Nevertheless this kind of comparisons will be interesting to test our formalism and it will be future work.



%

\section*{Acknowledgements}
I would like to thank Pietro Colangelo, Fulvia De Fazio, Claudio Corian\`o, Luca Trentadue and other organisers of the international workshop on QCD at Giovinazzo (Italy) for their hospitality. I also acknowledge Hisnag-nan Li for useful discussions and the financial support to participate in this workshop. This work was supported in part by the grant NSC-101-2112-M-001-006-MY3.

%

%
%

\end{document}